\documentclass[preprint,aps]{revtex4}

\usepackage{dcolumn}% Align table columns on decimal point
\usepackage{bm}% bold math

\begin{document}

%\preprint{}

\title{Effects of nonzero neutrino masses on black hole evaporation}

\author{Daniel Bambeck}
     \email{bambeck@physics.montana.edu}
\author{William A.\ Hiscock}
     \email{hiscock@montana.edu}
\affiliation{Department of Physics, Montana State University Bozeman,
Montana 59717}

\date{June 8, 2005}

\begin{abstract}
We study the consequences of nonzero neutrino masses for black
holes evaporating by the emission of Hawking radiation. We find
that the evolution of small, hot, black holes may be unaffected
(if neutrinos are Majorana particles), or may show an increase in
neutrino luminosity and a decrease in lifetime by up to a factor
of 1.85 (if neutrinos are Dirac particles). However, for
sufficiently large (e.g., stellar mass) black holes, neutrino
emission is largely or entirely suppressed, resulting in a
decrease in emitted power and an increase in lifetime by up to a
factor of 7.5.

\end{abstract}

%\pacs{Valid PACS appear here}

\maketitle

Since the discovery that black holes emit particles with a thermal
spectrum \cite{Hawking}, it has been understood that black holes
will have a finite lifetime due to this emission. Most analyses of
black hole evaporation have relied on the detailed studies
completed by Page \cite{Page1,Page2} of the flux of particles of
differing spins and masses emitted by Schwarzschild and Kerr black
holes.

The general picture that emerges is that any black hole in
isolation will lose any initial charge and angular momentum in the
early stages of evaporation, asymptotically approaching a
Schwarzschild-like state long before reaching microscopic size.
Specifically, any charged black hole with mass less than
approximately $10^5 M_{\odot}$ will rapidly lose any initial
electric charge \cite{Gibb,HW}. Furthermore, unless there are a
large number of undiscovered massless scalar fields in nature
\cite{TCH}, black holes will quickly spin down, with their
specific angular momentum $a=J/M$ decreasing at roughly twice the
rate their mass decreases \cite{Page2}. Hence, any black hole is
expected to asymptotically approach a Schwarzschild state,
nonrotating and uncharged, well before it completely evaporates.

The subsequent evaporation of the Schwarzschild-like black hole
was treated by Page \cite{Page1}. He found that 81\% of the power
emitted by a hole with $ M >> 10^{17}$ g is in the form of
neutrinos, assuming two species (electron and muon) of massless
neutrinos (87\% of the power would be in neutrinos with three
massless species). Today there is strong evidence that the three
known species of neutrino have nonzero masses. This evidence comes
from atmospheric \cite{Fuk1,Fuk2,Yasuda} and solar neutrino
experiments that imply neutrino oscillations are taking place
\cite{BKS}. In order for neutrinos to oscillate among species,
they must have mass. The details of the oscillations fix the
differences of the squares of the masses of the neutrinos, but not
their separate individual values. Oscillations occur because the
weak interaction eigenstates are not identical to the mass
eigenstates.

The purpose of this Letter is to examine the effect of nonzero
neutrino masses on the evolution of evaporating black holes. Since
so much of the emitted power for large black holes would be
massless neutrinos, noteworthy quantitative, if not qualitative,
differences in black hole evolution might be expected when
neutrino masses are taken into account.

We take a conservative viewpoint concerning unknown neutrino
physics, assuming, in particular, that there are only three
species of neutrino (i.\ e.\ , that no additional ``sterile''
neutrino species exist that do not participate in charged or
neutral current weak interactions). This assumption is favored by
the atmospheric neutrino oscillation data \cite{Fuk3}. We utilize
current experimental results to bound the possible changes in
black hole evaporation caused by nonzero neutrino masses within
these assumptions. Additional refinement is inevitable as the
newly recognized richness of neutrino physics is further explored.

The atmospheric data, with the assumption of three neutrino
species, is best fit by \cite{SKC}
\begin{equation}
    \Delta m_{13}^2 \simeq 2.4 \times 10^{-3} {\rm eV}^2 \; \; ,
\label{deltam2atm}
\end{equation}
while the solar neutrino oscillations combined with KamLAND
experimental data yield for the second squared mass difference
\cite{KamLAND}:
\begin{equation}
    \Delta m_{12}^2 \simeq 8 \times 10^{-5} {\rm eV}^2 \, \, ,
\label{deltam2slr}
\end{equation}
where (1,2,3) label the neutrino mass eigenstates.

These experimental results are compatible with two possible
spectra for the neutrino mass eigenstates, with either two closely
grouped low mass neutrinos and one higher (often termed the
``normal'' hierarchy), or vice versa (the ``inverted'' hierarchy).
The specific (as opposed to relative) values of the mass
eigenvalues are far less certain than the new knowledge of the
squared differences from neutrino oscillations. The lowest
possible set of masses is achieved by setting the lowest mass
eigenvalue to be infinitesimally larger than zero. The maximum
possible values for the neutrino masses are constrained by a
recent cosmological bound obtained from the $2$ Degree Field
Galaxy Redshift Survey \cite{Elgaroy},
\begin{equation}
    m_\nu^{\rm tot} < 1.8 \,{\rm  eV} \, \, ,
\label{masstot}
\end{equation}
where $m_\nu^{\rm tot}$ is the sum of the neutrino mass
eigenvalues, and assuming currently accepted values for the Hubble
constant and total matter density \cite{Spergel}.

An additional issue that arises once neutrinos are recognized to
be massive particles is whether the neutrino and antineutrino are
distinct particles. It may well be that neutrinos carry no
conserved quantum number that would distinguish particle from
antiparticle; if not, then neutrinos are Majorana particles. On
the other hand, if there is, say, a conserved lepton number, then
neutrinos are Dirac particles and neutrinos and antineutrinos are
distinct particles. In the massive Dirac case, neutrinos and
antineutrinos can each occur in two spin states, yielding twice as
many states as are available for massless neutrinos. If they are
Majorana particles, then the neutrino is its own antiparticle,
lepton number is not conserved, and there are only two states
available, as in the massless case. The Dirac or Majorana nature
of the neutrino is presently one of the most intriguing questions
in neutrino physics. In this Letter, we will describe the effects
on black hole evaporation of massive neutrinos of both types.

There are thus three key questions to which the answers are not
yet known, even within the conservative assumption of only three
neutrino species. First, what is the mass of the lightest
neutrino? Second, is the small (solar) mass splitting at high mass
or low mass? Third, are neutrinos Dirac or Majorana particles?

There are also, however, three crucial facts that are now
established by experiment and observation. First, neutrinos are
not massless particles. Second, even assuming the neutrino mass
spectrum favors low masses (the ``normal'' hierarchy), two of the
three neutrinos have masses of order at least $10^{-2}$ eV. Third,
based on cosmological observations, the heaviest neutrino has a
mass of at most about $2$ eV.

The new physics resulting from massive neutrinos can have
significant effects on black hole evaporation both for small, hot
black holes, and for large, cold black holes. We address three
questions in this Letter. First, how does the existence of massive
neutrinos change the predicted initial mass of a black hole that
is evaporating to zero mass today? Second, if massive neutrinos
are Dirac particles, by what factor can the lifetime of an
initially small, hot black hole be reduced? Third, for large, cold
black holes of astrophysical mass, by what factor will their
evaporation lifetimes be increased by massive neutrinos?

Following Page \cite{Page1}, we assume black holes emit particles
in each mode independently.  Therefore, the emission rate (and
emitted power) of each species of particle is directly
proportional to the number of modes allowed to that particle.  The
constants of proportionality depend on the spin and total angular
momentum of the particle, and were determined numerically by Page.

Because emission in all modes is independent, our work depends
primarily on counting available modes, without need to re-run
Page's code.  There are three ways in which our counting differs
from that of Page:  First, at the time of his paper, only two
species of neutrino, electron and muon, were known, whereas now
there are known to be three.  Secondly, emission in some or all
neutrino species may be suppressed, due to their mass.  Third,
neutrinos may have more spin states available than previously
thought.

The power emitted by an evaporating Schwarzschild black hole (and
hence, the rate of mass loss) is proportional to the inverse
square of the mass,
\begin{equation}
    {dM}/{dt} = -a/{M^2}
\label{massdiffeq}
\end{equation}
implying a lifetime of
\begin{equation}
    T = {M_0^3}/{3a}
\label{lifetimeeq}
\end{equation}
where $M$ is the mass of the hole at a given time, $M_0$ is the
initial mass, $a$ is a constant of proportionality which depends
on the particle species being emitted by the hole, and Planck
units are used. From Page's work \cite{Page1},
\begin{equation}
    a = 4.091 \times 10^{-5} n + 3.755 \times 10^{-5}
\label{adef}
\end{equation}
where $n$ is equal to the number of modes, considering particle
species and spin states, available in spin-1/2 particles, and the
final constant term is due to radiation in massless photons and
gravitons.

Over its lifetime, as a hole shrinks and becomes hotter, the
constant $a$ will increase in an almost stepwise fashion (hence
the justification in treating it to be constant over substantial
periods of  time), as the temperature of the hole passes
thresholds corresponding to the masses of various particles.
However, since a black hole spends most of its lifetime very near
its initial mass, it is a reasonable approximation to use the
initial mass to determine which species are available to be
emitted. As long as the initial temperature is not exceedingly
close to a mass threshold, the lifetime calculated using the
initial value of $a$ is a very good approximation to the precise
value, as originally noted by Page \cite{Page1}.

The first question we wish to address is what the initial mass of
a primordial black hole would be that is evaporating completely
(reaching zero mass) at the present time. It is easy to show that
such a black hole would be initially small enough, and therefore
hot enough, to emit all neutrinos, as well as electrons and
positrons, and of course massless photons and gravitons. If
neutrinos were massless and had definite helicity, as was
previously believed, then there would be six neutrino states
available (neutrinos and antineutrinos in each of the three
types), as well as four electron states (electrons and positrons,
with two spin states each) for a total of ten states available in
spin-1/2 particles. Assuming the black hole is the same age as the
universe, $1.37 \times 10^{10}$ years \cite{Spergel}, using
Eqs.(\ref{lifetimeeq}) and (\ref{adef}), this yields an initial
mass of $2.21 \times 10^{19}$ times the Planck mass, or $4.80
\times 10^{14} {\rm g}$. If neutrinos are massive Majorana
particles, then the number of neutrino modes available is the same
as for the case of massless neutrinos, and therefore the initial
mass is the same, since the initial black hole is more than hot
enough to produce all three types of massive neutrino.

However, if neutrinos are massive Dirac particles, then there are
a total of 12 neutrino states available (three types of neutrino
and (distinct) antineutrino, with two spin states each), for a
total of 16 spin-1/2 particle states.  This results in a higher
luminosity, so the initial mass must therefore be larger to last
the same amount of time.  Specifically, the initial mass must be
$2.55 \times 10^{19}$ times the Planck mass, or $5.56 \times
10^{14} {\rm g}$, some 16\% larger than the initial black hole
mass for massless neutrinos.

In either case, the initial temperature of the black hole is of
order 20 MeV, well above the masses of the electron and all
neutrino species, so the initial assumption of emission in
neutrinos and electrons is justified.

The possibility that massive neutrinos are Dirac particles leads
to the second scenario we wish to address: the (naively
surprising) result that massive neutrinos may cause small black
holes to evaporate more quickly than if neutrinos were massless.
The effect is most dramatic for black holes that are large and
cool enough initially so that electrons and positrons are not
copiously produced ($T << 0.511\; {\rm MeV}, M
>> 2 \times 10^{16} {\rm g}$), but are still small and hot enough
so that all three types of neutrinos are sure to be produced
initially ($T >> 2\; {\rm eV}, M << 5 \times 10^{21} {\rm g}$). In
this case, the existence of additional neutrino states that the
black hole can emit into can decrease the lifetime of such a hole
by as much as 46\%, a factor of 1.85.

The third situation in which neutrino masses can be important in
black hole evaporation is for large black holes whose lifetimes
greatly exceed the age of the universe. For such large black
holes, a critical issue is whether they are initially hot enough
to emit substantial numbers of neutrinos.

Since a black hole radiates thermally with a temperature
proportional to its surface gravity, emission of particles with a
rest mass higher than the surface temperature will be
exponentially suppressed.  In particular, this implies that a hole
can only emit significant quantities of particles with rest mass
less than the temperature of the hole, or, in Planck (natural)
units, $m << 1/8\pi M$, where $m$ is the mass of the particle
emitted. For a stellar mass black hole of three solar masses, this
temperature corresponds to $1.8 \times 10^{-12} {\rm eV}$.

However, the smallest possible neutrino mass splitting is $\Delta
m^2 \simeq 8 \times 10^{-5} {\rm eV}^2$ , which implies that at
least two neutrinos have masses greater than $10^{-2} {\rm eV}$,
and certainly all three {\it could} have masses greater than
$10^{-12} {\rm eV}$. This leads, then, to three possibilities
which must be considered for an astrophysical mass black hole:  No
neutrinos initially emitted, one type of Majorana neutrino
initially emitted, or one type of Dirac neutrino initially
emitted. Again, since any black hole spends most of its lifetime
near its initial mass, only the number of species emitted at the
initial temperature is relevant to the approximate lifetime.

In the first case, the lightest neutrino has a mass greater than
the temperature of the hole, and therefore no neutrinos (or any
other spin-1/2 particles) will be initially emitted. Since no
neutrinos are emitted, it does not matter whether they are
Majorana or Dirac particles. This leads to a lifetime of $8877
M_0^3$, or $1.157 \times 10^{67} ({M_0}/{M_{\odot}})^3$ years.

In the second case, we assume the neutrino is a Majorana particle,
and that the lightest neutrino has a mass less than the initial
temperature of the black hole. In this case, there are two spin
states of that light neutrino that will be initially emitted. The
lifetime of the hole would then be $2792 M_0^3$, or $3.638 \times
10^{66} ({M_0}/{M_{\odot}})^3$ years.

In the third case, we assume the neutrino is a Dirac particle, and
that the lightest neutrino has a mass less than the initial
temperature of the black hole. In this case, there are four states
available: Neutrino or antineutrino, with two spin states each.
The lifetime is then $1657 M_0^3$, or $2.159 \times 10^{66}
({M_0}/{M_{\odot}})^3$ years.

For comparison, if neutrinos are considered to be massless
particles with definite helicity, then there are six neutrino
states (particle and antiparticle, of each of three types), that
will be emitted initially, and the life-span is therefore $1178
M_0^3$, or $1.535 \times 10^{66} ({M_0}/{M_{\odot}})^3$ years.

Thus, in comparison with the massless neutrino case, the existence
of neutrino mass may increase the evaporation lifetime of large,
astrophysical mass black holes by as much as a factor of $7.5$.

\begin{acknowledgements}
This work was supported in part by NSF Grant No. PHY-0098787.
\end{acknowledgements}
% ==========================================================

% ==========================================================

\end{document}